\documentclass[aps,prl,reprint,
amsmath,amssymb,
groupedaddress,
showpacs]{revtex4-1}
\usepackage{color}
\usepackage{braket}
\usepackage{mathrsfs}
\usepackage{graphicx}
\usepackage{graphics}
\usepackage{amssymb, latexsym}
\usepackage{amsmath}
\usepackage{bm}
\usepackage{color}
\usepackage{ulem}
\usepackage{fancybox}
\usepackage{xspace}
\usepackage{amsfonts}
\usepackage{multirow}
\usepackage[section]{placeins}
\usepackage{dcolumn}
\usepackage{amsthm}
\usepackage{setspace}
\usepackage{dsfont}
\usepackage{here}

\def\nn{\nonumber}

\def\0{\bm 0}

\begin{document}

\title{Topological Invariant for Bosonic Bogoliubov-de Gennes Systems with Disorder}

\author{Yutaka Akagi\thanks{E-mail address: E-mail address: yutaka-akagi@g.ecc.u-tokyo.ac.jp}
}

\affiliation{Department of Physics, Graduate School of Science, The University of Tokyo, Hongo, Tokyo 113-0033, Japan
}

\date{\today}

\begin{abstract}
Using the method of noncommutative geometry, we define a topological invariant in disordered bosonic Bogoliubov-de Gennes systems, 
which possess a unique mathematical property---non-Hermiticity.
To demonstrate the validity of the definition, we investigate a disordered artificial spin ice model in two dimensions numerically. 
In the clean limit, we clarify that the topological index perfectly coincides with the Chern number. 
We also show that the topological index is robust against disorder. 
The formula provides the topological index $n_{\rm Ch}=1$ in the magnon Hall regime and $n_{\rm Ch}=0$ in a trivial localized one. 
We also show by example that our method can be extended to other symmetry classes. Our results pave the way for further studies on topological bosonic systems with disorder.
\end{abstract}

\pacs{}

\maketitle

In recent decades, the classification of topological insulators/superconductors and the related phenomena have been studied intensively~\cite{Schnyder08, Kitaev09, Ryu10, Hasan10, Qi11}. 
Nontrivial topology in materials usually results in the appearance of robust edge states, which are also expected to provide potential applications~\cite{Stern13, Yue17, Zeng20}.
Such nontrivial phases are characterized by certain topological invariants. 
A well-known example is the Chern number, which corresponds one-to-one with the number of chiral edge states in a quantum Hall system, known as the bulk-boundary correspondence~\cite{Hatsugai93a, Hatsugai93b}.

It has been realized that bosonic systems also exhibit topological phenomena, such as the topological (thermal) Hall effect.
Examples range from systems of photons~\cite{Onoda04, Hosten08, Raghu08, Haldane08, Wang09a, Ben-Abdallah16},
to systems of phonons~\cite{Strohm05, Sheng06a, Inyushkin07, Kagan08, Wang09b, Zhang10, Qin12, Mori14, Sugii17}, 
magnons~\cite{Fujimoto09, Katsura10, Onose10, Matsumoto11a, Matsumoto11b, Ideue12, Shindou13a, Shindou13b, Matsumoto14, Mook14, Chisnell15, Hirschberger15, Kim16, Owerre16a, Owerre16b, Han_Lee17, Murakami_Okamoto17, Nakata17, Wang17, Wang18, Seshadri18, Kawano19a, Kim19}, 
and triplons~\cite{Rumhanyi15, McClarty17}.
Among the various bosonic systems, research on topological aspects of magnonic systems has increased rapidly. 
Recently, it has also been found that magnonic systems provide symmetry-protected topological phases, 
for instance, the ${\mathbb Z}_2$ topological phases~\cite{ Zyuzin16, Kondo19a, Kondo19b} and higher-order topological phases~\cite{Hirosawa20, Mook20}. 
As a difference from fermionic systems, bosonic Bogoliubov-de Gennes (BdG)-type systems, which contain pairing terms,
possess non-Hermicity intrinsically due to the bosonic statistics.
Therefore, the topological classification of Hermitian systems is not directly applicable to bosonic BdG systems~\cite{Kawabata19, Kondo20}.

Topological phases are expected to be robust against disorder owing to their topological nature.
However, it is not obvious how to define topological invariants in the presence of disorder due to broken translational symmetry.
Recently, it was found that the approach of noncommutative geometry~\cite{Avron94a, Avron94b, Aizenman98} provides 
mathematically rigorous representations of topological invariants for the ten Altland-Zirnbauer classes~\cite{Katsura18}.
The numerical method for calculating the invariants has been established as well~\cite{Akagi17}.
Various other expressions within the symmetry classes have also been defined,
such as the Chern number in real space~\cite{Kitaev06, Prodan09, Prodan10} 
and the Bott index~\cite{Bellissard94, Hastings10, Loring10, Loring15, Huang18a, Huang18b}.

However, a numerical determination of the topological index for bosonic BdG systems with disorder is still very challenging.
Roughly speaking, two methods have been proposed so far for this problem: 
the $C^*$-algebraic approach~\cite{Peano18} and the consctruction of the real-space Bott index~\cite{Wang20}.
In this paper, we propose an alternative and highly efficient method to numerically calculate the topological index for magnon Hall systems with disorder.
The method is based on the noncommutative index theory~\cite{Avron94a, Avron94b, Aizenman98, Akagi17} 
and has an advantage in that the extension of the method to other symmetry classes is straightforward. 
Some examples are given at the end of the paper.

Our method is demonstrated for the artificial spin ice model with disorder, which exhibits the magnon Hall effect~\cite{Xu16}.
As in fermionic systems, the noncommutative index in the clean limit perfectly coincides with the conventional Chern number~\cite{Shindou13b}
corresponding to the number of chiral edge states.
We also show that the topological index is robust against disorder, and the index $n_{\rm Ch}=1$ ($0$) characterizes 
the disordered topological (a trivial localized) magnon phase.

We begin with the bosonic BdG Hamiltonian
\begin{align}
&\mathcal{H}_{\rm BdG}=\frac{1}{2}\sum_{\bm{k}}\bm{\beta}^{\dagger} H_{\rm BdG}\bm{\beta}, \label{eq:Ham}\\
&\bm{\beta}^\dagger = [ b^\dagger_1, \cdots, b^\dagger_{\mathscr{N}}, b_1, \cdots, b_{\mathscr{N}}].\label{eq:op_phi}
\end{align}
Here, $b_{j}^{\dagger}$ $(b_{j})$ denotes boson creation (annihilation) operator with label $j$. 
The subscript $\mathscr{N}$ is the total number of degrees of freedom in a system
(in this paper, the $\mathscr{N}$ is the total number of sites).
The matrix $H_{\rm BdG}$ is written as
\begin{align}
H_{\rm BdG}=\left(
\begin{array}{cc}
h & \Delta \\
\Delta^{*}&h^{*} \\
\end{array} 
\right).
\end{align}
Since the $2\mathscr{N} \times 2\mathscr{N}$ matrix $H_{\rm BdG}$ is Hermitian, the $\mathscr{N} \times \mathscr{N}$ matrices $h$ and $\Delta$ satisfy $h^{\dagger}=h$ 
and $\Delta^{T}=\Delta$, respectively.
Here, ${O^*}$, ${O^{\dagger}}$, and ${O^T}$ are the complex conjugate, adjoint, and transposition of a matrix $O$, respectively.
The components of the operator $\bm{\beta}$ satisfy the commutation relation $[\beta_{i},\beta_{j}^{\dagger}]=(\Sigma_{z})_{ij}$.
Here, $\Sigma_{z}$ is defined as a tensor product $\Sigma_{a}:=\sigma_a \otimes  1_\mathscr{N}$, 
where $\sigma_a$ $(a=x, y, z)$ is the $a$-component of the Pauli matrix acting on the particle-hole space and $1_\mathscr{N}$ is the $\mathscr{N} \times \mathscr{N}$ identity matrix.

To hold the bosonic commutation relation, the Hamiltonian $H_{\rm BdG}$ is diagonalized by a para-unitary matrix $T$, 
which satisfies $T^{\dagger} \Sigma_z T = \Sigma_z$.
Solving the eigenvalue problem $\Sigma_z H_{\rm BdG}{\bm v}_j = E_j {\bm v}_j$, 
we obtain the eigenvalues $E_j$ of $H_{\rm BdG}$ and the para-unitary matrix (eigenvectors ${\bm v}_j$) automatically:
\begin{align}
T^{-1} \Sigma_z H_{\rm BdG} T&= {\rm diag}[ E_1, \cdots, E_{\mathscr{N}}, -E_1, \cdots, -E_{\mathscr{N}}],
\end{align}
where the para-unitary matrix is given by
\begin{align}
T= [ {\bm v}_1, \cdots, {\bm v}_{\mathscr{N}}, {\bm v}_{\mathscr{N}+1}, \cdots, {\bm v}_{2\mathscr{N}}]. 
\label{eq:paraunitary}
\end{align}
Since the effective Hamiltonian $\Sigma_z H_{\rm BdG}$ is non-Hermitian, 
the inner-product for bosonic wave functions is modified as
\begin{align}
\langle \! \langle\bm{\phi},\bm{\psi}\rangle \! \rangle=\bm{\phi}^{\dagger}\Sigma_z \bm{\psi},
\end{align}
where $\bm{\phi}$ and $\bm{\psi}$ are $2\mathscr{N}$-dimensional complex vectors~\cite{Lein19}.

Next, for defining the topological invariant, we introduce a ``Fermi" projection $P_{\rm B}$ for bosonic BdG systems 
described by
\begin{align}
P_{\rm B} &= \frac{1}{2 \pi i}\oint _{\mathcal{C}} (z-\Sigma_z \mathcal{H}_{\rm BdG})^{-1}dz \nonumber \\
&= \sum_{E_n \le E_{\rm B}} (\Sigma_z)_{nn} \bm{v}_n \bm{v}_n^{\dagger} \Sigma_z,
  \label{eq:PB_finite}
\end{align}
where the contour $\mathcal{C}$ encloses all the spectrum of $\Sigma_z \mathcal{H}_{\rm BdG}$ below the ``Fermi" energy for bosons $E_{\rm B}$.
As in fermionic systems, the $P_{\rm B}$ is the projection onto the states whose energies are less than the fictitious Fermi energy $E_{\rm B}$. 
The element of Eq.~(\ref{eq:PB_finite}) $P_n:=(\Sigma_z)_{nn} \bm{v}_n \bm{v}_n^{\dagger} \Sigma_z$ satisfies 
$P_m P_n = \delta_{mn}P_n$ (i.e., $P_{\rm B}^2 = P_{\rm B}$)
and $\sum_{n=1}^{2\mathscr{N}} P_n = 1_{2\mathscr{N}}$,
and thus the effective Hamiltonian is rewritten as $\Sigma_z H_{\rm BdG} = \sum_n E_n P_n$ (spectral decomposition).
We note that the projection $P_n$ is no longer Hermitian as the operator satisfies $\Sigma_z P_n^{\dagger} \Sigma_z = P_n$.

\begin{figure}[t]
\begin{center}
\includegraphics[width=1.0\columnwidth]{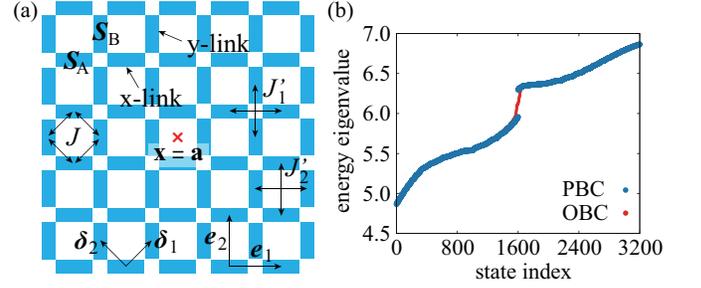}
\end{center}
\caption{(Color online).
(a) Artificial spin ice model on a square lattice with the nearest (next nearest) neighbor dipole coupling $J$ ($J^{\prime}_{1}$ and $J^{\prime}_2$).
A(B)-sublattice spins ${\bm S}_{{\bm i}\in A}$ (${\bm S}_{{\bm i}\in B}$) are located on the center of the $x$($y$)-link.
We define vectors in Eq.~(\ref{boson-H}) as $e_1=(1,0)$, $e_2=(0,1)$, $\delta_1 = (1/2,1/2)$, and $\delta_2 = (-1/2,1/2)$.
The location of ${\bm a}$ in Eq.~(\ref{eq:Dirac_op}) is chosen to be the center of the lattice. 
(b) Magnon energy spectrum of Eq.~(\ref{boson-H}) in the clean limit with PBC and OBC. 
The topological edge state shown in red traverses the energy gap with PBC between 5.96 and 6.30.
The parameters used in (b) are 
\mbox{$J=1.0$}, \mbox{$J_1^{\prime}=J_2^{\prime}=0.35$}, \mbox{$D=2.0$}, \mbox{$H^z=20.0$}, \mbox{$S=1.0$} and the system size is \mbox{$L^2=3200$}.
}
\label{Fig:model_spec}
\end{figure}

As in fermionic systems, we also introduce the Dirac operator 
$\mathcal{{\tilde D}}_{\bm a}$ for bosonic BdG systems as 
\begin{equation}
\mathcal{{\tilde D}}_{\bm a} := \sigma_z \otimes \mathcal{D}_{\bm a}({\bm x}), \,\,
\mathcal{D}_{\bm a}({\bm x}) := \frac{x + i y - (a_x + i a_y)}{| x + i y - (a_x + i a_y) |}.
\label{eq:Dirac_op}
\end{equation}
Here, ${\bm x} =$ \mbox{$(x,  y)$} is the position vector of lattice points, 
and ${\bm a} =$ \mbox{$(a_x, a_y) \neq {\bm x} $} is a vector of a flux point [see Fig.~\ref{Fig:model_spec}(a)].

Next, we define the topological index for tight-binding models of bosonic BdG systems with disorder. 
The difference of two projections is defined as \cite{Avron94a, Avron94b}
\begin{equation}
A:=P_{\rm B} - \mathcal{{\tilde D}}_{\bm a}^{\dagger} P_{\rm B} \mathcal{{\tilde D}}_{\bm a}
\label{eq:A_op}
\end{equation}
where $\mathcal{{\tilde D}}_{\bm a}^{\dagger}$ is the adjoint of the Dirac operator $\mathcal{{\tilde D}}_{\bm a}$. 
Suppose that $A^{2n+1}$ for $n \in \mathbb{N}$ is trace class. 
Then, the topological index $n_{\rm Ch}$ is defined as
\begin{eqnarray}
  n_{\rm Ch} := {\rm dim} \; {\rm ker}\; [A - 1] - {\rm dim} \; {\rm ker}\; [A + 1],
  \label{eq:index}
\end{eqnarray}
where ${\rm dim} \; {\rm ker}\; O$ stands for the dimension of the kernel of an operator $O$.
We note two properties of the operator $A$: 
(i) $\|A\| < 1$ since the operator $A$ consists of the difference between two projections.
(ii) The eigenvalues of the non-Hermitian operator $A$ are real 
because of the relation $\Sigma_z A^{\dagger} \Sigma_z = A$, which is derived from $\Sigma_z P_n^{\dagger} \Sigma_z = P_n$.

We here show the underlying supersymmetric structure of the operator $A$. 
According to Refs.~\cite{Avron94a, Avron94b}, a pair operator of $A$ is defined as
\begin{eqnarray}
B:=1-P_{\rm B} - \mathcal{{\tilde D}}_{\bm a}^{\dagger} P_{\rm B} \mathcal{{\tilde D}}_{\bm a}.
  \label{op:B}
\end{eqnarray}
The operators $A$ and $B$ satisfy the following two relations:
\begin{eqnarray}
AB+BA=0 \; \; \; {\rm and} \; \; \; A^2+B^2=1.
  \label{op:AB}
\end{eqnarray}
Let $\varphi_1$ be an eigenvector of $A$ with an eigenvalue $\lambda$, i.e., $A \varphi_1 = \lambda \varphi_1$.
Then, the anti-commutation relation of Eq.~(\ref{op:AB}) yields
\begin{eqnarray}
AB\varphi_1 = -BA\varphi_1 = - \lambda B \varphi_1.
  \label{op:pairing}
\end{eqnarray}
The second relation of Eq.~(\ref{op:AB}) produces
\begin{align}
&B^2 \varphi_1 = (1 - \lambda^2) \varphi_1 \; \; \; {\rm and} \nonumber \\
&\langle \! \langle B \varphi_1, B \varphi_1 \rangle \! \rangle = (1-\lambda^2) \langle \! \langle \varphi_1,\varphi_1 \rangle \! \rangle,
  \label{norm_phi2}
\end{align}
where we used $\Sigma_z B^{\dagger} \Sigma_z = B$, which is also derived from $\Sigma_z P_n^{\dagger} \Sigma_z = P_n$.
Equations~(\ref{op:pairing}) and (\ref{norm_phi2}) imply that $B$ is an invertible map from an eigenvector with an eigenvalue $\lambda$ to that with $-\lambda$, except for $\lambda=0$, $\pm 1$.
In other words, $\lambda$ and $- \lambda$ for $0< |\lambda | <1$ always come in pairs 
while $\lambda =\pm 1$ are isolated points of the spectrum of $A$~\cite{comment_SUSY}.

\begin{figure}[t]
\begin{center}
\includegraphics[width=0.8\columnwidth]{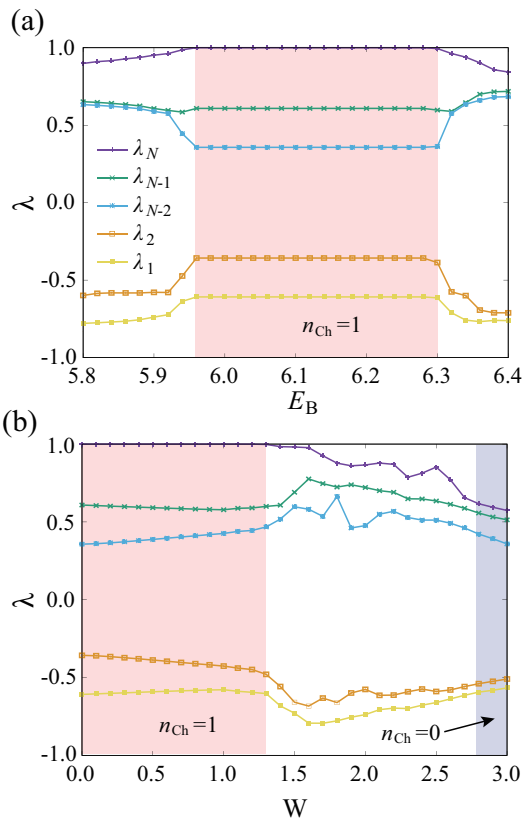}
\end{center}
\caption{(Color online).
(a) Fictitious Fermi level $E_{\rm B}$ and (b) disorder strength $W$ dependences of the eigenvalues 
$\lambda_1$, $\lambda_2$, $\lambda_{N-2}$, $\lambda_{N-1}$, and $\lambda_{N}$ of the operator $A$. 
(a) and (b) correspond to cross-sections of Fig.~\ref{Fig:PD_spin-ice} for \mbox{$W=0.0$} and \mbox{$E_{\rm B}=6.2$}, respectively.
The red (blue) colored region is the magnon Hall (trivial localized) regime characterized by $n_{\rm Ch} = 1$ ($n_{\rm Ch} = 0$).
The parameters used in (a) and (b) are 
\mbox{$J=1.0$}, \mbox{$J_1^{\prime}=J_2^{\prime}=0.35$}, \mbox{$D=2.0$}, \mbox{$H^z=20.0$}, \mbox{$S=1.0$} and the system size is \mbox{$L^2=3200$}.
}
\label{Fig:cross-section}
\end{figure}

Let us consider the robustness of the index against a perturbation $\delta  \mathcal{H}_{\rm BdG}$ as discussed in Refs.~\cite{Katsura16a, Katsura18}.
The projection for the total Hamiltonian $\mathcal{H}_{\rm BdG}' := \mathcal{H}_{\rm BdG} + \delta \mathcal{H}_{\rm BdG}$ is written as
\begin{eqnarray}
P_{\rm B}' = \frac{1}{2 \pi i}\oint _{\mathcal{C}} (z-\Sigma_z \mathcal{H}_{\rm BdG}')^{-1}dz,
  \label{PB_pert}
\end{eqnarray}
and then we introduce $A' :=P_{\rm B}' - \mathcal{{\tilde D}}_{\bm a}^{\dagger} P_{\rm B}' \mathcal{{\tilde D}}_{\bm a}$ as the $A$ operator for $\mathcal{H}_{\rm BdG}'$.
We here consider the following two assumptions:  
(i) The range of the hopping integrals of $\delta  \mathcal{H}_{\rm BdG}$ is finite, i.e., the norm $\| \delta  \mathcal{H}_{\rm BdG} \|$ is finite.
(ii) The fictitious Fermi level for bosons still lies in the spectral gap of $\mathcal{H}_{\rm BdG}'$.
Using the contour integral, we write the difference between $P_{\rm B}'$ and $P_{\rm B}$ as
\begin{align}
P_{\rm B}' - & P_{\rm B} = \nonumber \\
 \frac{1}{2 \pi i} &\oint _{\mathcal{C}} \frac{1}{z-\Sigma_z \mathcal{H}_{\rm BdG}'} \Sigma_z \delta \mathcal{H}_{\rm BdG} \frac{1}{z-\Sigma_z \mathcal{H}_{\rm BdG}}dz.
  \label{dif_PB}
\end{align}
Because of the spectral gap assumption, the norm of the right-hand side can be bounded by the norm $\| \Sigma_z \delta  \mathcal{H}_{\rm BdG} \|$. 
Then, the difference $A' - A = (P_{\rm B}' -  P_{\rm B}) - \mathcal{{\tilde D}}_{\bm a}^{\dagger} (P_{\rm B}' - P_{\rm B}) \mathcal{{\tilde D}}_{\bm a}$ is bounded by the norm, 
and thus the operator $A$ is continuous with respect to $\| \Sigma_z \delta  \mathcal{H}_{\rm BdG} \|$. 
Using the Weyl's inequality for the real spectrum~\cite{Bhatia07},
we obtain the desired results that the nonzero eigenvalue $\lambda$ 
of the operator $A$ is continuous with respect to $\| \Sigma_z \delta  \mathcal{H}_{\rm BdG} \|$
[see the spectrum in red regions of Fig.~\ref{Fig:cross-section} as we discuss later in detail].
Consequently, the topological index given by Eq.~(\ref{eq:index}) is robust against perturbations.

As a demonstration, we compute the topological index $n_{\rm Ch}$ of the disordered artificial spin ice model on a square lattice, 
which exhibits the magnon Hall effect [refer to the original spin Hamiltonian in Ref.~\cite{Xu16}].
In this model, 
$A$($B$)-sublattice spin on $x$($y$)-link of the square lattice has an easy-axis anisotropy $D_{\bm j}$ along
the $x$($y$)-direction due to a magnetic shape anisotropy~\cite{wang06,budrikis10,budrikis12,iacocca16} 
[see Fig.~\ref{Fig:model_spec}(a)].
We here consider a fully polarized state by the out-of-plane strong magnetic field $H_{\bm j}^z$.
By applying the Holstein-Primakoff transformation, the magnon Hamiltonian of the model is written as 
\begin{align}
H_{\rm BdG} &:=  H_{\rm on} + H_{\rm nn} + H_{\rm nnn} \nn \\
H_{\rm on} \equiv & \frac{1}{2} \sum_{{\bm j} \in A} \big\{(D + d_{\bm j})S (-{b^{\dagger}_{{\bm j}}}^2
- b^{\dagger}_{\bm j} b_{\bm j}) + (H^z + h_{\bm j}) b^{\dagger}_{\bm j} b_{\bm j}\big\}  \nn \\
& \hspace{-1.0cm} + \frac{1}{2} \sum_{{\bm j} \in B} \big\{(D + d_{\bm j}) S ({b^{\dagger}_{{\bm j}}}^2
- b^{\dagger}_{\bm j} b_{\bm j}) + (H^z + h_{{\bm j}}) b^{\dagger}_{\bm j} b_{\bm j}\big\} + {\rm h.c.} \nn
\end{align}
\begin{align}
H_{\rm nn} \equiv & \sum_{m=1,2} \sum_{{\bm j} \in A}
\sum_{{\bm i}={\bm j}\pm \delta_{m},{\bm i} \in B}  \nn \\
& \hspace{-1.2cm}  \frac{J S}{2}
\big(  - b^{\dagger}_{\bm j} b_{\bm i}   - b^{\dagger}_{\bm i}b_{\bm i}
 - b^{\dagger}_{\bm j} b_{\bm j} + 3i (-1)^m  b^{\dagger}_{\bm j} b^{\dagger}_{\bm i} + {\rm h.c.} \big) \nn \\
H_{\rm nnn} \equiv & \sum_{\alpha=A,B} \sum_{m=1,2}
\sum_{{\bm j} \in \alpha} \sum_{{\bm i}={\bm j}\pm e_{m}}  \nn \\
&\hspace{-1.2cm}  \frac{J^{\prime}_{\alpha,m} S}{4}
\big(  - b^{\dagger}_{\bm j} b_{\bm i} - b^{\dagger}_{\bm i}b_{\bm i}
 - b^{\dagger}_{\bm j} b_{\bm j} + 3 (-1)^m  b^{\dagger}_{\bm j} b^{\dagger}_{\bm i} + {\rm h.c.} \big), \label{boson-H}
\end{align}
where $b^{\dagger}_{\bm j}$ ($ b_{\bm j}$) is the magnon creation (annihilation) operator at site ${\bm j}$, and $S$ is the spin magnitude.
The coupling constant $J$ and $J^{\prime}_{\alpha,m}$ denote the nearest and the next nearest dipolar interactions, 
respectively with $J^{\prime}_{A,1}=J^{\prime}_{B,2} = J^{\prime}_1$ and $J^{\prime}_{A,2}=J^{\prime}_{B,1} = J^{\prime}_2$. 
As shown in Fig.~\ref{Fig:model_spec}(a), $e_{1}$ and $e_{2}$ are the primitive lattice vectors of the square lattice and $\delta_{m}=[e_1-(-1)^{m} e_2]/2$ ($m=1,2$). 
The easy-axis anisotropy and magnetic field include randomness as
$D_{\bm j}=D+d_{\bm j}$ and $H^z_{\bm j}=H^z+h_{\bm j}$ where $D$ and $H^z$ are uniform components.
The randomness $d_{\bm j}$ and $h_{\bm j}$ are uniformly distributed within $[-W/2, W/2]$ with a positive parameter $W$.

Figure~\ref{Fig:model_spec}(b) shows the magnon energy spectrum, in ascending order, of the model in the absence of disorder.
The blue/red dots are the energy spectrum under periodic/open boundary condition (PBC/OBC).
As shown in the figure, the topological edge state (OBC) traverses the energy gap between 5.96 and 6.30 (PBC). 
This is because the Chern number of the bottom (top) band is $+1$ ($-1$), which is computed in terms of the Bloch wave-function in crystal momentum space.

To evaluate the topological index, we plot $E_{\rm B}$ and $W$ dependences of the eigenvalues $\lambda_i$ ($i=1,2, N-2, N-1, N$) 
of the operator $A$ as shown in Fig.~\ref{Fig:cross-section}
[see Ref.~\cite{Akagi17} for detailed numerical implementation]. 
Here, we write $\lambda_i$ for the {\it i}-th eigenvalue of the operator $A$ in ascending order, i.e., $(-1 \le)$ $\lambda_1 \le \lambda_2 \le \cdots \le \lambda_N$ $(\le 1)$.
As shown in the red regions of Fig.~\ref{Fig:cross-section}(a) and (b),
\mbox{$\lambda_N \simeq 1$}, \mbox{$\lambda_1 \neq -1$}, and \mbox{$\lambda_N - \lambda_{N-1} \neq 0$}, 
and hence $n_{\rm Ch} = 1$.
The region of $n_{\rm Ch} = 1$ for $W=0.0$ (absence of disorder) in Fig.~\ref{Fig:cross-section}(a) perfectly coincides with that with the Chern number $+1$. 
As discussed above, one can see the pairing structure as $\lambda_{N-1} \simeq -\lambda_1$ and $\lambda_{N-2} \simeq -\lambda_2$ in the red regions.
One can also see, in the red regions, that the spectrum is continuous. 
On the other hand, in the white regions, the spectrum fluctuates, and the pairing structure is broken.
We note that our method is not effective because the spectral or mobility gap is expected to vanish in the white regions~\cite{Akagi17}.

The blue region of Fig.~\ref{Fig:cross-section}(b) clearly shows the reconstruction of the pairing structure as $\lambda_{N} \simeq -\lambda_1$ and $\lambda_{N-1} \simeq -\lambda_2$. 
This implies that in the blue region, magnons are localized by disorder.
In the localized regime, the eigenvalue $\lambda_N$ is significantly different from unity, and thus $n_{\rm Ch} = 0$.
From the observation of Fig.~\ref{Fig:cross-section}(a) and (b), we find that
the largest (smallest) eigenvalue $\lambda_N$ ($\lambda_1$) and the difference between $\lambda_N$ and $|\lambda_1|$, i.e., \mbox{$\lambda_N + \lambda_1$},
give us information about the determination of the topological index $n_{\rm Ch}$.

\begin{figure}[t]
\begin{center}
\includegraphics[width=1.0\columnwidth]{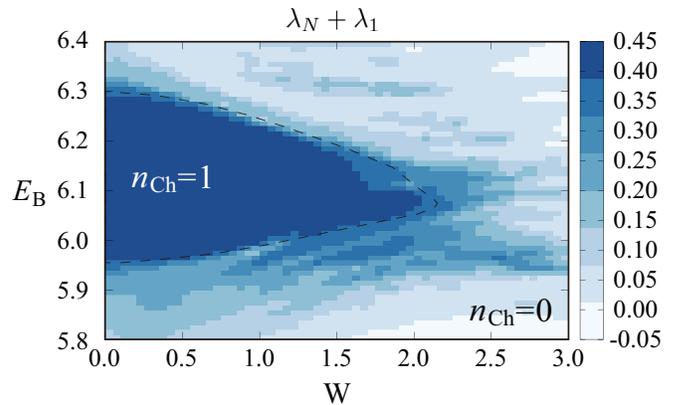}
\end{center}
\caption{(Color online).
Dependence of \mbox{$\lambda_N + \lambda_1$} with the topological index $n_{\rm Ch}$ in the artificial spin ice model 
as a function of the disorder strength $W$ and the fictitious Fermi energy $E_{\rm B}$. 
The largest eigenvalue $\lambda_N$ is larger than $0.995$ in the blue region inside of the dotted line. 
The parameters are the same as in Fig.~\ref{Fig:cross-section}.
}
\label{Fig:PD_spin-ice}
\end{figure}

Figure~\ref{Fig:PD_spin-ice} represents \mbox{$\lambda_N + \lambda_1$} as a function of the strength of disorder $W$ and the fictitious Fermi level $E_{\rm B}$. 
In the blue region enclosed by the dotted line of Fig.~\ref{Fig:PD_spin-ice}, $\lambda_N > 0.995$ and \mbox{$\lambda_N + \lambda_1 \simeq 0.4$}, and hence $n_{\rm Ch} = 1$.
In the localized regime of \mbox{$\lambda_N + \lambda_1 \simeq 0$}, 
the topological index $n_{\rm Ch}$ is zero as discussed in the Fig.~\ref{Fig:cross-section}(b).
However, as in fermionic systems~\cite{Akagi17}, it is difficult to identify the ``metallic" phase, in which the spectral or mobility gap vanishes, only by the method of noncommutative geometry.
It is beyond our scope of this paper to determine clear phase boundaries.

Summary: 
We have provided a new method for numerically calculating the topological index of magnon Hall systems with disorder. 
Introducing the ``Fermi" projection for bosons $P_{\rm B}$, we have defined the topological index $n_{\rm Ch}$ 
determined by the discrete spectrum of the operator $A$ with a supersymmetric structure. 
To check the validity of our method, we have computed the topological index for the artificial spin ice model with disorder~\cite{Xu16}.
As in fermionic systems~\cite{Akagi17}, we have clarified that in the clean limit, 
the region with the topological index $n_{\rm Ch}=1$ perfectly coincides with 
that with the Chern number $+1$ corresponding to the number of chiral edge states.
We have also shown that the index $n_{\rm Ch}=1$ ($0$) characterizes the disordered topological (trivial localized) magnon phase.

Finally, we remark that the generalization of our method to models in other symmetry classes is straightforward.
For instance, the ${\mathbb Z}_2$ index $\nu$ in bosonic BdG systems with pseudo-time-reversal symmetry in two and three dimensions~\cite{Kondo19a, Kondo19b, Kondo20} 
is defined as
\begin{eqnarray}
  \nu = {\rm dim} \; {\rm ker} [A - 1] \; \; {\rm modulo} \; 2,
  \label{eq:Z2-index}
\end{eqnarray}
as in fermionic systems~\cite{Katsura16a, Katsura18, Akagi17}.
Here, in two dimensions, the definition of the operator $A$ is the same as Eq.~(\ref{eq:A_op}).
In three dimensions, the operator $A$ is given by
\begin{eqnarray}
A := P_B - {\tilde D}_a P_B {\tilde D}_a,
\label{eq:A_op_3D}
\end{eqnarray}
where the Dirac operator ${\tilde D}_a$ is defined as
\begin{eqnarray}
{\tilde D}_{\bm a}
 := \sigma_z \otimes D_{\bm a}({\bm x}), \,\,
D_{\bm a}({\bm x}) := \frac{1}{|{\bm x} - {\bm a}|} ({\bm x} - {\bm a}) \cdot \mbox{\boldmath $\sigma$}.
\label{eq:Da_3D}
\end{eqnarray}
Here, 
${\bm x} =$ \mbox{$(x, y, z)$} 
is the position operator of lattices and 
${\bm a} =$ \mbox{$(a_x, a_y, a_z) \neq {\bm x}$} 
is a vector of flux point as in Fig.~\ref{Fig:model_spec}(a). 
The vector $\mbox{\boldmath $\sigma$}$ is defined by 
$\mbox{\boldmath $\sigma$}=\mbox{$(\sigma_x, \sigma_y, \sigma_z)$}$.
We will report the ${\mathbb Z}_2$ index $\nu$ in detail in a future article.

\acknowledgements
{
The author thanks Hosho Katsura for helpful discussions and for sharing his insights on the subject.
The author also acknowledges useful discussions with Ken Shiozaki.
This work was supported by JSPS KAKENHI Grants No. JP17K14352, JP20K14411, 
JSPS Grant-in-Aid for Scientific Research on Innovative Areas ``Topological Materials Science" (KAKENHI Grant No. JP18H04220), and 
``Quantum Liquid Crystals" (KAKENHI Grant No. JP20H05154). 
The author also thanks the Okinawa Institute of Science and Technology Graduate University for the use of the facilities, Sango and Deigo clusters.
}


\begin{thebibliography}{101}
\bibitem{comment_SUSY}
The supersymmetry structure produces Eq.~(\ref{eq:index}) as
${\rm Tr}A^{2n+1} =\sum_{\lambda > 0} \lambda^{2n+1} n_{\lambda} + (- \lambda)^{2n+1} n_{-\lambda} =n_{1} - n_{-1} = n_{\rm Ch}$,
where $n_{\lambda}$ denotes the multiplicity of $\lambda$. 
Here, we used Lidskii's theorem.~\cite{Simon79}
\end{thebibliography}
\end{document}